\documentclass[pra,aps,amsmath,amssymb,twocolumn]{revtex4}

\usepackage{graphicx}
\usepackage{dcolumn}
\usepackage{bm}

\begin{document}

\title{A combined experimental and theoretical study on realizing and using laser controlled torsion of molecules}

\author{C.~B.~Madsen}
\email[Corresponding author: ]{cbm@phys.au.dk}\affiliation{Lundbeck Foundation Theoretical Center for Quantum
System Research, Department of Physics and Astronomy, Aarhus
University, 8000 Aarhus C, Denmark}

\author{L.~B.~Madsen}
\affiliation{Lundbeck Foundation Theoretical Center for Quantum
System Research, Department of Physics and Astronomy, Aarhus
University, 8000 Aarhus C, Denmark}

\author{S.~S.~Viftrup}
\affiliation{Department of Chemistry, Aarhus University, 8000 Aarhus
C, Denmark}

\author{M.~P.~Johansson}
\affiliation{Department of Chemistry, Aarhus University, 8000 Aarhus
C, Denmark}

\author{T.~B.~Poulsen}
\affiliation{Department of Chemistry, Aarhus University, 8000 Aarhus
C, Denmark}

\author{L.~Holmegaard}
\affiliation{Department of Chemistry, Aarhus University, 8000 Aarhus
C, Denmark}

\author{V.~Kumarappan}
\affiliation{Department of Chemistry, Aarhus University, 8000 Aarhus
C, Denmark}

\author{K.~A.~J{\o}rgensen}
\affiliation{Department of Chemistry, Aarhus University, 8000 Aarhus
C, Denmark}

\author{H.~Stapelfeldt}
\email[Corresponding author: ]{henriks@chem.au.dk}
\affiliation{Department of Chemistry and Interdisciplinary
Nanoscience Center (iNANO), Aarhus University, 8000 Aarhus C,
Denmark}

\begin{abstract}
It is demonstrated that strong laser pulses can introduce torsional
motion in the axially chiral molecule
3,5-diflouro-3',5'-dibromo-biphenyl (DFDBrBPh). A nanosecond laser
pulse spatially aligns the stereogenic carbon-carbon (C-C) bond axis
allowing a perpendicularly polarized, intense femtosecond pulse to
initiate torsional motion accompanied by a rotation about the fixed
axis. We monitor the induced motion by femtosecond time-resolved
Coulomb explosion imaging. Our theoretical analysis corroborates the
experimental findings and on the basis of these results we discuss
future applications of laser induced torsion, viz., time-resolved
studies of de-racemization and laser controlled molecular junctions
based on molecules with torsion.
\end{abstract}
\maketitle

\section{Introduction}

Gaining control over the external and the internal degrees of
freedom of molecules is of great interest for a range of areas in
molecular science.  One approach, dating as far back as the
1920s~\cite{Kallmann:ZP:1921,Stern:ZP:1927}, relies on the
application of inhomogenous (static or low frequency) magnetic and
electric fields. This has proven most useful for controlling the
full three-dimensional motion of molecules, including control of the
velocity distribution, bringing molecules to a stand still, storing
them for extended periods of times and trapping them in a confined
volume (for a recent review see~\cite{meijer:natphys:2008}). An
alternative and complementary approach relies on the use of
non-resonant non-ionizing laser fields typically supplied by pulsed
lasers. Many studies during the last decade have shown that strong
non-resonant laser fields can effectively manipulate the external
degrees of freedom of isolated gas phase molecules. The manipulation
results from laser-induced forces and torques due to the interaction
between the induced dipole moment and the laser field itself.
Examples of manipulation include deflection~\cite{stapelfeldt:1997}, focusing~\cite{chung:jcp:2001} and
slowing~\cite{Fulton:NP:2006} of molecules through the dependence of
the non-resonant polarizability interaction on the intensity
distribution in a laser focus. Likewise, the dependence of the
induced dipole interaction on molecular orientation has proven
highly useful for controlling the rotation of a variety of
molecules~\cite{stapelfeldt:2003:rmp}. In particular, the spatial
orientation of molecules can be sharply confined with respect to
axes that are fixed in the laboratory.

Molecular manipulation by induced dipole forces extends beyond the
external degrees of freedom and has also been demonstrated for the
internal degrees of freedom such as vibrational
motion~\cite{Niikura:prl:2003} in molecular hydrogen. Notably, the
electrical field from laser pulses can modify energy potential
barriers such that photoinduced bond breakage of a small linear
molecule is guided to yield a desired final
product~\cite{Sussman:science:2006,Sussman:pra:2005}.

In the case of larger molecules control of the lowest frequency
vibrational modes attracts special interest since some of these
modes correspond to motion along well-defined reaction coordinates
separating two conformational minima (conformers). Although many
molecules contain large number of conformers it is often just two
conformers that dominate important chemical properties, for instance
chirality. A particularly important example is found in axially
chiral molecules ~\cite{Eliel_Wilen_1994,Bringmann:2005} such as
biaryl systems. In these molecules rotation about a single
stereogenic carbon-carbon (C-C) bond axis changes the molecule from
one enantiomer into the opposite enantiomer (mirror image).
Recently, we demonstrated that the laser-induced nonresonant
polarizability interaction can also be used to influence the
internal rotation of an axially chiral molecule around the
stereogenic C-C bond axis~\cite{madsen:2009:prl}. In particular, we
showed that by fixing the C-C bond axis of a substituted biphenyl
molecule in space, using laser induced alignment by a long laser
pulse, it was possible to initiate torsional motion of the two
phenyl rings by a short laser pulse polarized perpendicular to the
fixed axis. The purpose of the present work is to extend  our recent
paper~\cite{madsen:2009:prl}. In particular, we discuss our
theoretical modelling in detail.

\section{Outline of the strategy for laser controlled torsion}
 A ns laser pulse aligns
\cite{stapelfeldt:2003:rmp} the stereogenic axis of
3,5-diflouro-3',5'-dibromo-biphenyl (DFDBrBPh) along its
polarization. This enables a much shorter fs pulse, which we will refer to as the kick pulse, to both initiate
torsional motion and set the molecule into rotation around the fixed
axis. The internal as well as external rotational motion is
monitored by fs time-resolved Coulomb explosion imaging.

\begin{figure}
    \centering
   \includegraphics[width=\columnwidth]{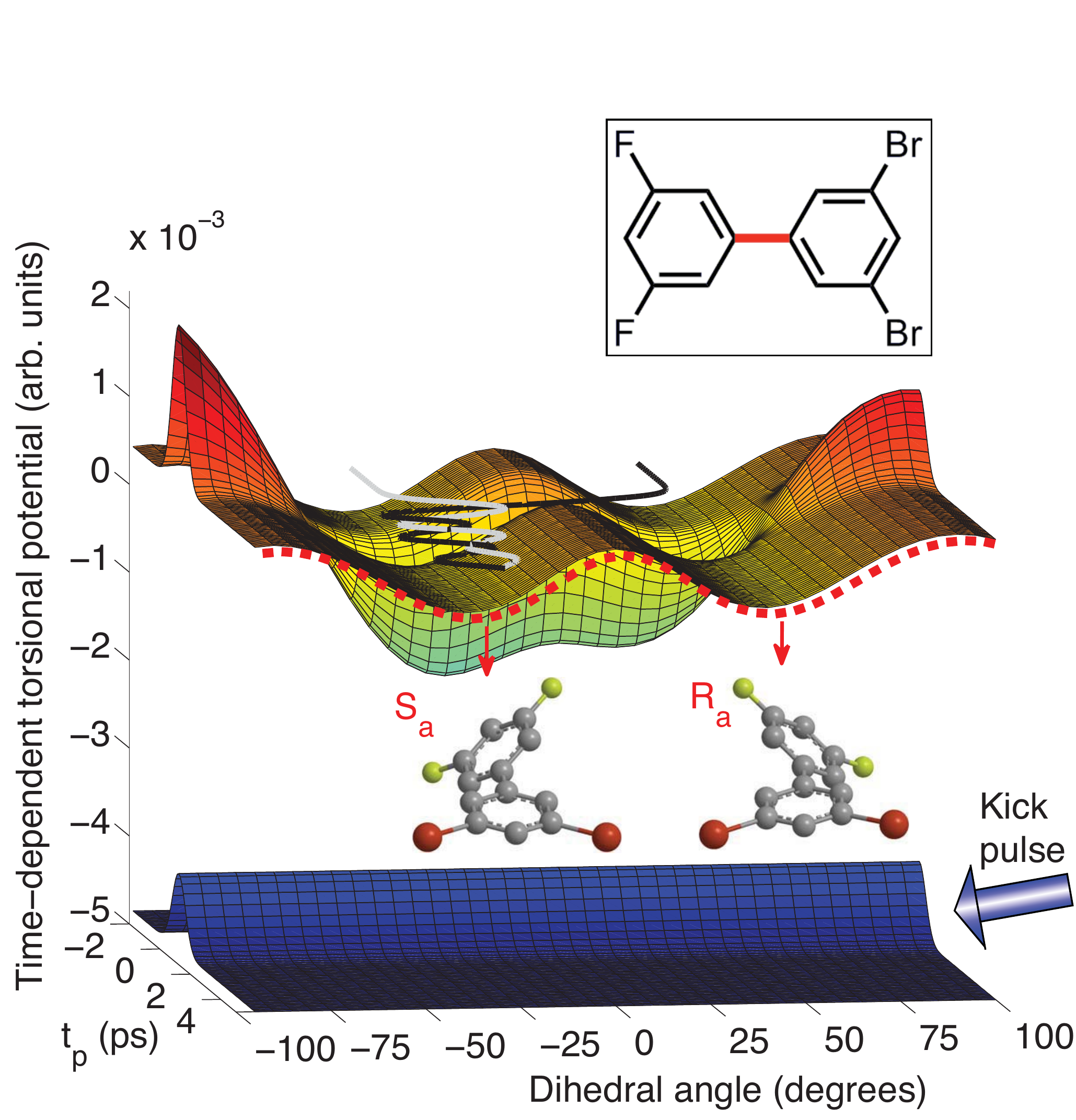}
 \caption{(Color online) To illustrate the principle of laser-induced torsion and time-resolved de-racemization we show
the calculated kick pulse induced time-dependent torsional potential
as a function of time measured with respect to the center of the
pulse and dihedral angle $\phi_d$ between the Br- and F-phenyl
planes. The asymmetry in the potential is obtained by orienting the
molecules (here with the Br-phenyl plane out of the paper), 3D
aligning them, and by polarizing the kick pulse at an angle of
13$^\circ$ with respect to the second most polarizable axis (SMPA)
(see Fig.~\ref{fig2}). The red dotted curve at large times
illustrates the laser-free time-independent torsional potential. For
the S$_a$ enantiomer, starting out with
$\langle\phi_d\rangle_i=-39^\circ$, the time varying potential
induces an oscillatory motion (grey curve) corresponding to torsion
confined within the initial well. By contrast, due to the induced
asymmetry between the two wells, the initial R$_a$ enantiomer is
traversing the central torsional barrier, and ends up as an S$_a$
enantiomer undergoing internal rotation (black curve). The kick
pulse has an intensity of $1.2\times10^{13}$ W$/$cm$^2$ and a
duration of 1.0 ps (FWHM). The torsional motion may be monitored by
fs time-resolved Coulomb explosion imaging. The inset shows a model
of the DFDBrBPh molecule with the sterogenic axis marked by red
(grey).}
  \label{fig1}
\end{figure}

The inset of Fig.~\ref{fig1} shows a model of the DFDBrBPh molecule
with the stereogenic axis marked by red. According to our quantum
chemical calculations, the laser-free torsional potential,
illustrated by the red dotted curve in Fig.~\ref{fig1}, has minima
at dihedral angles of $\langle \phi_d \rangle =\pm 39^\circ$
corresponding to the R$_a$ and S$_a$ enantiomeric forms, and where
the dihedral angle is the angle between the two phenyl rings. The
twisted equilibrium shape is characteristic of biphenyl compounds.
The traditional view is that this non-planarity results from
competition between stabilization via conjugation of the
$\pi$-orbital systems and steric repulsion between the
ortho-positioned atoms \cite{Grein_2002}. We might mention that
alternative explanations have recently been discussed
\cite{C.Matta:CEJ.2003,J.Poater:CEJ.2006,L.Pacios:SC.2007}. Our
strategy for controlling $\phi_d$ relies on a transient modification
of the field-free potential curve by a fs kick pulse
(Fig.~\ref{fig1}). The modification, caused by the nonresonant
polarizability
interaction~\cite{stapelfeldt:2003:rmp,Sussman:science:2006,Sussman:pra:2006}
converts the initial stationary quantum states, localized near the
minima of the torsional potential, into vibrational wave packets,
i.e., coherent superpositions of several quantum states. The
temporal evolution of these wave packets gives rise to
time-dependent torsional motion, and the grey and the black curves
in Fig.~\ref{fig1} show the calculated expectation values of
$\phi_d$ for a particular set of laser parameters. The stereogenic
C-C bond axis is fixed in the laboratory frame by adiabatic
alignment utilizing a moderately intense, linearly polarized ns
laser pulse \cite{stapelfeldt:2003:rmp,kumarappan:2006:jcp}. The ns
pulse is intense enough to keep the C-C axis tightly confined, yet
weak enough to modify the torsional potential only slightly. The fs
kick pulse is applied with its polarization perpendicular to the
aligned C-C bond axis to ensure primarily influence on torsion while
avoiding other possible vibrational motion.

The DFDBrBPh compound is prepared as a racemate and the experiment is carried out on isolated molecules at
rotational temperatures of a few Kelvin. Under these conditions we have equal numbers of molecules
initially localized in the $-39^\circ$ or 39$^\circ$
conformation (see Fig.~\ref{fig1}) and no thermally induced transitions
between the two conformations occur. The intense probe pulse, sent
at time $t_p$ with respect to the kick pulse, removes several
electrons from the molecules, thereby triggering Coulomb explosion
into ionic fragments. In particular, the Br$^+$ and F$^+$ fragment
ions recoil in the planes defined by the Br- and F-phenyl rings. By
recording the velocities of both ion species with two-dimensional
ion imaging \cite{kumarappan:2006:jcp}, we determine the instantaneous
orientation of each of the two phenyl planes at the time of the
probe pulse.

\section{Theoretical methods}
\subsection{Modelling the dynamics of the molecule}
\label{sec:AlignmentEffectiveHam} 
\begin{figure}
    \centering
    \includegraphics[width=\columnwidth]{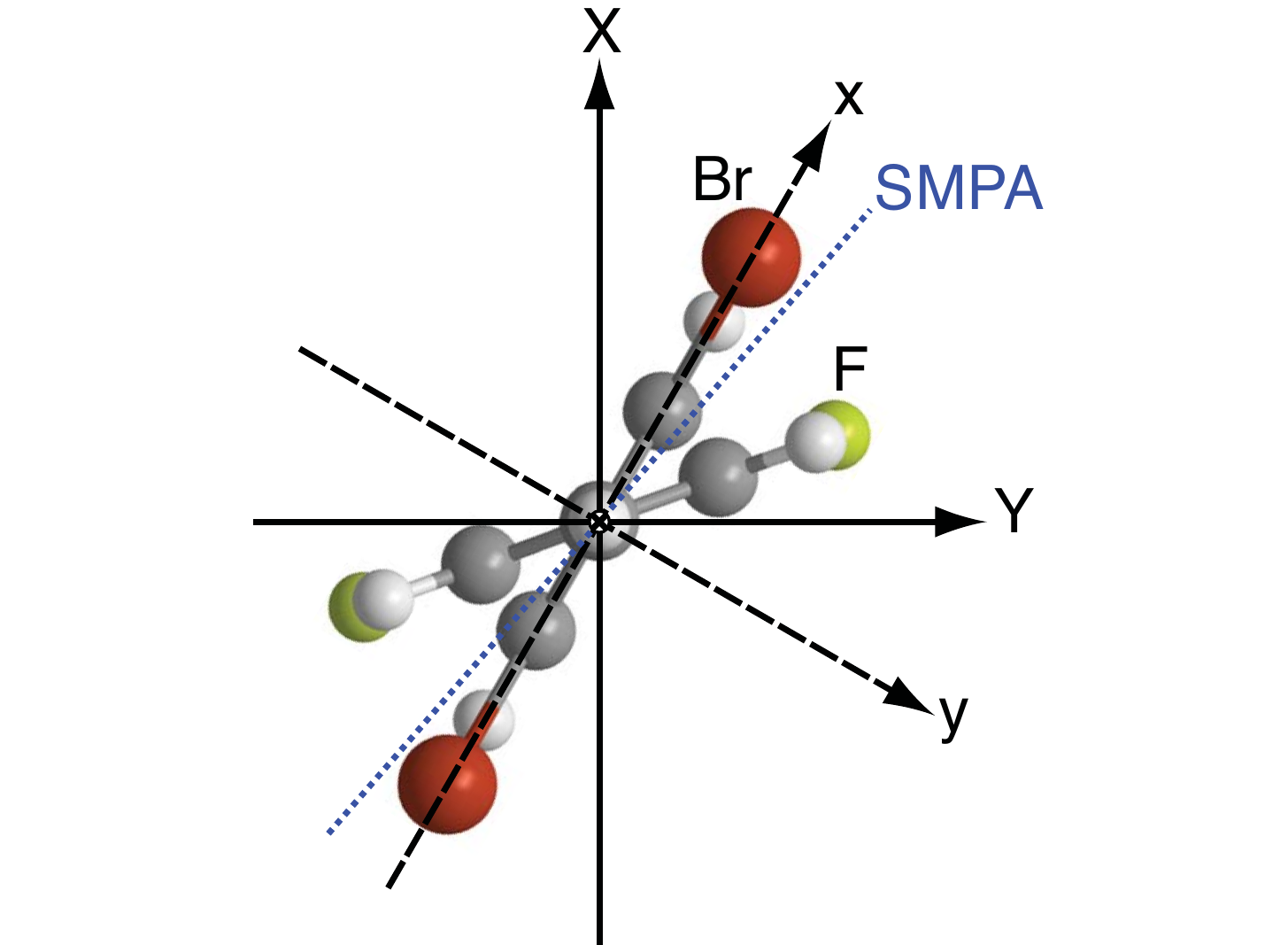}
  \caption{Model of DFDBrBPh along with the molecular fixed (MF)
 $xyz$ and the laboratory  fixed $XYZ$ coordinate axes. The dotted line indicates the second most polarizable axis (SMPA), which is located 11$^\circ$ from the Br-phenyl ring and 28$^\circ$ from the F-phenyl ring.}
  \label{fig2}
\end{figure}

To account for the motion of the
nuclei of the DFDBrBPh molecule we make use of two sets of
coordinate systems (see Fig.~\ref{fig2}): A molecular fixed (MF) frame attached to the
molecule and a laboratory fixed (LF) frame specified by the
lasers. The MF coordinates are chosen with the $z$ axis along the
stereogenic axis pointing from the phenyl ring with the bromines towards
the phenyl ring with the flourines, and the $x$ axis is chosen along the
phenyl ring with the bromines. The LF coordinates are chosen with the $Z$ axis along
the polarization direction of the ns pulse and the $X$ axis along
the kick pulse polarization direction. In agreement with the
experimental observations (cf. Sec.~\ref{sec:results}), we assume that the stereogenic axis of
the DFDBrBPh molecule is aligned along the $Z$ axis.
Also, we
neglect all normal modes, except the lowest one, which
corresponds to torsion (see Sec.~\ref{subsec:QMcalc} for details on quantum
chemical calculations, justifying, e.g., this assumption). In this simplified situation and
in the absence of the kick pulse the task is reduced to describing
the coupled rotations of the two (rigid) phenyl rings of the molecule as
given in the LF frame by the Hamiltonian [Atomic units,
$m_e=e=a_0=\hbar=1$, are used throughout unless indicated
otherwise]
\begin{equation}\label{eq:HamBP}
H_\text{mol}=-\frac{1}{2I_\text{Br}}\frac{\partial^2}{\partial\phi_\text{Br}^2}-\frac{1}{2I_\text{F}}\frac{\partial^2}{\partial\phi_\text{F}^2}+V_\text{tor}(\phi_{Br}-\phi_{F}),
\end{equation}
where $\phi_i,i=\text{Br},\text{F}$, is the angle of the $i$ phenyl ring
with respect to the kick pulse polarization axis, $I_i$ the moment
of inertia for rotation of the $i$ phenyl ring around the stereogenic axis ($I_\text{Br}=8911925$, $I_\text{F}=1864705$) and
$V_\text{tor}(\phi_\text{Br}-\phi_\text{F})$ is the laser-free
torsional potential as obtained from quantum chemical calculations (Fig.~3). By changing the coordinates to the dihedral angle
$\phi_d=\phi_\text{Br}-\phi_\text{F}$ between the two phenyl rings and the
weighted azimuthal angle
$\Phi=(1-\eta)\phi_\text{Br}+\eta\phi_\text{F}$,
cha\-rac\-te\-ri\-zing the rotation of the molecule, with
$\eta=I_\text{F}/(I_\text{F}+I_\text{Br})$, we obtain
\begin{align}\label{eq:HamBPmod}
H_\text{mol}&=\left(-\frac{1}{2I}\frac{\partial^2}{\partial\Phi^2}\right)+\left(-
\frac{1}{2I_\text{rel}}\frac{\partial^2}{\partial\phi_d^2}+
V_\text{tor}(\phi_d)\right)\nonumber\\&=H_\Phi+H_{\phi_d}.
\end{align}
Here $I=I_\text{Br}+I_\text{F}$ is the total moment of inertia of
the molecule for rotation around the stereogenic axis and $I_\text{rel}=
I_\text{Br}I_\text{F}/I$ is a relative moment of inertia for the two
phenyl rings around the axis. A full rotation of either phenyl ring leaves us
with the same molecule implying $2\pi$-periodic boundary
conditions of the eigenfunctions of $H_{\text{mol}}$ from
Eq.~\eqref{eq:HamBP}, i.e., $\psi(\phi_\text{Br}+2\pi
m,\phi_\text{F}+2\pi n)=\psi(\phi_\text{Br},\phi_\text{F})$, with
$m$ and $n$ integers. We shall assume that this property also applies to $\Phi$ and $\phi_d$, so that we simply need to consider
eigenfunctions $\tilde{\psi}(\Phi,\phi_d)=\xi(\Phi)\chi(\phi_d)$ of
Eq.~\eqref{eq:HamBPmod} that separates into rotation of
the molecule as described by the $2\pi$-periodic function
$\xi(\Phi)$ and torsion accounted for by the $2\pi$-periodic
function $\chi(\phi_d)$. [Rigorously, the bounds for $\Phi$ depend
on $\phi_d$~\cite{Hoki2001}. At the time scales of interest,
however, a molecule generally only makes a small fraction of a full rotation so that the effect of $\Phi$-$\phi_d$ coupling is negligible.] The
separation is physically motivated by considering the energy scales
related to rotation and torsion. The energy scale
of the prior is given by $\hbar^2/(2I)=1.3~\mu$eV. For torsion, on the other hand, the relevant energy is determined by
the torsional potential, and a harmonic approximation of the
potential near the minimum at $39^\circ$ yields a frequency
corresponding to the energy $3.1$~meV. Using that the
period of motion is of the order Planck's constant divided by the
energy, we therefore see that the molecule rotates ($\Phi$ change) on a nanosecond time scale, whereas the torsion
($\phi_d$ change) is of picosecond duration.

\begin{figure}
    \centering
    \includegraphics[width=\columnwidth]{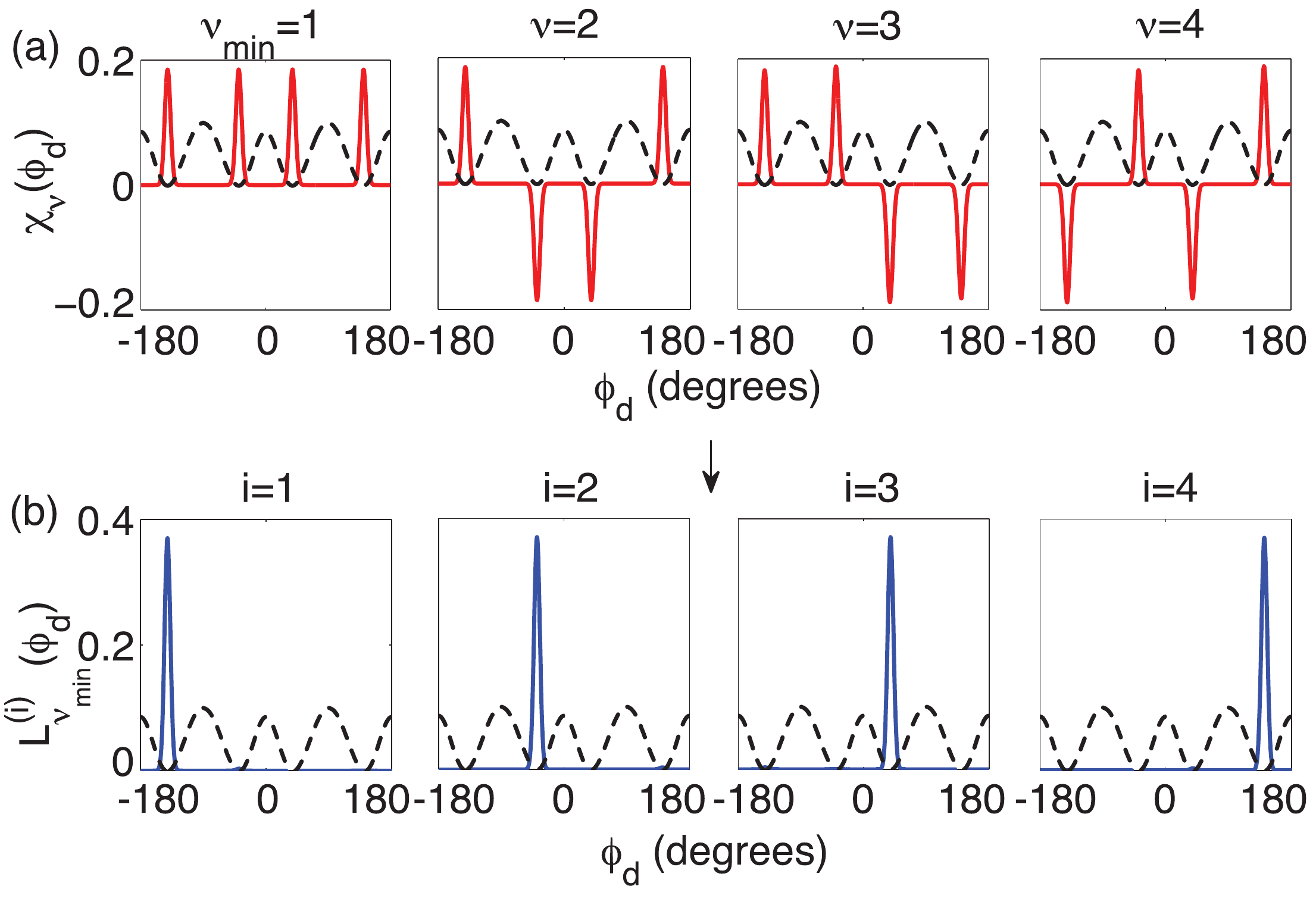}
  \caption{The field-free torsional states. (a) The four first (almost) degenerate energy eigenstates lying $1.71$ meV above the minimum of the torsional potential. From linear combinations of these we obtain the corresponding four localized states shown in (b). The torsional potential is indicated with a dashed line.}
  \label{fig3}
\end{figure}

The small energy separation of the rotational levels compared to that of the torsional levels implies that many rotational states will be occupied in thermal equillibrium. Consequently, we expect the $\Phi$ dynamics to behave classically (see also Fig.~\ref{fig7}), and we only treat the torsional dynamics quantum mechanically. The torsion is
described by the stationary Schr\"{o}dinger equation
\begin{equation}\label{eq:fftor}
H_{\phi_d}\vert\chi_\nu\rangle=E_\nu\vert\chi_\nu\rangle,
\end{equation}
with $\nu=1,2,\ldots$ denoting the energy eigenstates. To solve this equation
 we expand the Hamiltonian onto an
orthonormal basis of $2\pi$-periodic functions, which we diagonalize. The eigenstates below the torsional
barriers exhibit an almost exact fourfold degeneracy, and linear combinations of the approximately degenerate
states will also be stable at the time scales of the
experiment. In particular, we can produce such stable states that are localized in
the wells of the torsional potential. We will denote these states by
$\vert L_{\nu_\text{min}}^{(i)}\rangle$, with $i=1,2,3,4$ and
${\nu_\text{min}}$ the smallest $\nu$ of the four degenerate states.
In Fig.~\ref{fig3} we show the first four energy
eigenstates and the corresponding localized states along with the torsional potential.
The experiment is generally carried out on a gas of molecules in a
mixed state given by the density matrix
\begin{align}\label{eq:densDFDBrBP}
\rho(t_0)&= \sum_{\nu_\text{min}}\sum_{i=1}^4
P_{\nu_\text{min}}^{(i)}\vert L_{\nu_\text{min}}^{(i)}\rangle\langle
L_{\nu_\text{min}}^{(i)}\vert+\sum_\nu P_\nu \vert
\chi_\nu\rangle\langle \chi_\nu\vert,
\end{align}
with the $P$'s being weight factors that sum to unity, and where the
latter sum includes the states above the torsional barrier. The kick pulse introduces dynamics that we determine from the
alignment potential created by the pulse. By definition
$\bm{F}(t)=F_0(t)\hat{\bm{X}}$ for the kick pulse, and we use a
Gaussian envelope
$F_0(t)=F_0\exp\{-\ln(2)[t/(\tau_\text{FWHM}/2)]^2/2\}$
corresponding to an intensity full width at half maximum (FWHM) of
$\tau_\text{FWHM}$. We calculate the polarizability tensor in
the MF frame as a function of the dihedral angle (see
Table~\ref{tab:table1}) and consequently transform into the LF
frame by the application of directional cosine matrices~\cite{Zare}.
We then arrive at the interaction
\begin{align}\label{eq:kickpot}
V_\text{kick}(\Phi,\phi_d,t)&=-\frac{1}{4}F_0^2(t)[\alpha_\text{xx}(\phi_d)\cos^2(\Phi+\eta\phi_d)\nonumber\\
&+\alpha_\text{yy}(\phi_d)\sin^2(\Phi+\eta\phi_d)\nonumber\\
&-2\alpha_\text{xy}(\phi_d)\cos(\Phi+\eta\phi_d)\sin(\Phi+\eta\phi_d)].
\end{align}

\begin{table*}
\centering \caption{\label{tab:table1}The table lists the relevant
polarizability components, $\alpha_\text{ij}$, of DFDBrBPh in the MF
frame as a function of the dihedral angle, $\phi_d$. The components
are $\pi$-periodic and fulfill
$\alpha_\text{xx}(\phi_\text{d})=\alpha_\text{xx}(\pi-\phi_\text{d})$,
$\alpha_\text{yy}(\phi_\text{d})=\alpha_\text{yy}(\pi-\phi_\text{d})$
and
$\alpha_\text{xy}(\phi_\text{d})=-\alpha_\text{xy}(\pi-\phi_\text{d})$.
Also, $\alpha_\text{yx}=\alpha_\text{xy}$.}
\begin{ruledtabular}
\begin{tabular}{ll|rrrrrrr}
$\phi_\text{d}$ &(degrees)  & 0 & 15 & 30& 45 & 60 & 75 & 90\\
\hline
$\alpha_\text{xx}$ &(atomic units) & 217.694 & 215.590 & 209.463 & 200.975 & 192.431 & 186.240 & 184.048 \\
$\alpha_\text{yy}$ &(atomic units) & 92.352 & 95.201 & 102.634 & 112.658 & 122.693 & 130.015 & 132.639 \\
$\alpha_\text{xy}$ &(atomic units) & 0.000 & -9.488 & -16.360 &
-18.810 & -16.225 & -9.295 & 0.000\\
\end{tabular}
\end{ruledtabular}
\end{table*}

For the succeeding analysis it is helpful to note a few
qualitative features of the potential and we refer to Fig.~\ref{fig2} for graphical guidance. For a fixed dihedral angle
the potential is minimal when the second most polarizable axis (SMPA), which is located 11$^\circ$ from the Br-phenyl ring and 28$^\circ$ from the F-phenyl ring, is
parallel to the $X$ axis. We will denote this by the
$\parallel$-geometry. Conversely, the potential is maximal if the
molecule is rotated $90^\circ$ from the $\parallel$-geometry, and we
will denote this by the $\perp$-geometry.
Next, in the $\parallel$-geometry the potential favors a reduction
of the dihedral angle, whereas an increase of the dihedral angle is
resulting from the $\perp$-geometry. Hence, the overall effect of
the kick pulse will be to align the molecules into the
$\parallel$-geometry and drive a change of the dihedral angle.

We proceed with the quantitative analysis of the field-induced
dynamics. The state $\vert\chi_\nu\rangle$ is no longer an
eigenstate of the torsion when the kick pulse is applied,
but will develop into a wave packet
\begin{equation}\label{fieldtorstate}
\vert\chi_\nu\rangle\to\vert\chi_\nu^\Phi(t)\rangle=\sum_{\nu'}c_{\nu'}^\Phi(t)e^{-iE_{\nu'}(t-t_0)}\vert\chi_{\nu'}\rangle
\end{equation}
with the coefficients determined by
\begin{equation}\label{eq:coefftor}
\dot{c}_{\nu'}^\Phi(t)=-i\sum_\nu
c_\nu^\Phi(t)e^{-i(E_\nu-E_{\nu'})(t-t_0)}\langle\chi_{\nu'}\vert
V_\text{kick}(\Phi,t)\vert\chi_\nu\rangle.
\end{equation}
Once we find these new states of torsion the expectation value of any operator $O$ can be evaluated by tracing the product of the density matrix with the operator,
i.e., $\langle O(t)\rangle=\text{Tr}[\rho(t)O]$. In particular, we
need the expectation value of the dihedral angle and of the
kick potential from Eq.~\eqref{eq:kickpot}. From the latter we
obtain the torque, which causes the molecule to rotate into the
$\parallel$-geometry. If the molecule lies at an angle $\Phi$ it
is exposed to a torque $-\partial\langle
V_\text{kick}(\Phi,t)\rangle/\partial\Phi$ directed along the $Z$
axis and hence achieves an angular acceleration given by
\begin{equation}\label{eq:Phiacc}
I\ddot{\Phi}=-\frac{\partial\langle
V_\text{kick}(\Phi,t)\rangle}{\partial\Phi}.
\end{equation}
Along with the initial conditions given at
Eq.~\eqref{eq:densDFDBrBP}, the Eqs.~\eqref{eq:coefftor}
and~\eqref{eq:Phiacc} provide a set of coupled differential
equations that may be integrated to obtain the coordinates $\Phi(t)$
and $\langle\phi_d(t)\rangle$ at time $t$. Rather than solving these
coupled equations, we, however, assume that the angle $\Phi$ has the
constant value $\Phi_0$ during the short time interval of the kick
pulse, and we integrate Eq.~\eqref{eq:Phiacc} twice to
arrive at
\begin{equation}\label{eq:Phit}
\Phi(t)=\left.\Phi_0-t\frac{1}{I}\left(\frac{\partial}{\partial\Phi}\int_{-\infty}^\infty
dt' \langle
V_\text{kick}(\Phi,t')\rangle\right)\right\arrowvert_{\Phi=\Phi_0}.
\end{equation}
Consistently, we solve Eq.~\eqref{eq:coefftor} with $\Phi$ fixed
at $\Phi_0$.

For the present work we start out with a distribution of values for $\Phi_0$ in the
interval $[-\pi,\pi]$ along with an initial density matrix,
$\rho(t_0)$ and we propagate each member of the distribution according to
Eqs.~\eqref{eq:coefftor} and~\eqref{eq:Phit}. This procedure yields an
ensemble of $\Phi(t)$'s and $\langle \phi_d(t)\rangle$'s and we finally employ the
relations  $\phi_\text{Br}(t)=\Phi(t)+\eta\langle \phi_d(t)\rangle$ and
$\phi_\text{F}(t)=\Phi(t)-(1-\eta)\langle \phi_d(t)\rangle$ to find the
angular distributions of bromines and flourines.

\subsection{Quantum chemical calculations}\label{subsec:QMcalc}

To determine the field-free torsional potential we obtain the geometries of DFDBrBPh in various conformations at the density
functional theory level~\cite{P.Hohenberg:PR.1964, W.Kohn:PR.1965},
with B3LYP, Becke's three-parameter hybrid exchange
functional~\cite{A.Becke:JCP.1993b} in connection with the
Lee-Yang-Parr correlation \cite{C.Lee:PRB.1988} functional. We model the
correlation of the uniform electron gas with the
Vosko-Wilk-Nusair VWN5 formulation~\cite{S.Vosko:CJP.1980}, and we use doubly
polarized triple-zeta quality basis-sets,
TZVPP~\cite{A.Schafer:JCP.1994}. This level of theory has
been shown to be suitable for the study of
biphenyls~\cite{mikael08}. To scan the potential
energy surface of the relative torsional angle of the phenyl rings we optimize all coordinates except the torsional angle.

We do all electronic structure calculations with the
\textsc{Turbomole} 5.10 program suite~\cite{R.Ahlrichs:CPL.1989}. In
Table~\ref{tab:table1} we give the dynamic polarizabilities computed
at a laser wavelength of 1064 nm. We note that the polarisabilities are quite stable with respect to choice of laser frequency. Using a value of 800 nm changes the individual components only minutely (maximally by 2\%). The differences between static and dynamic polarisabilities are equally small. We calculate the vibrational frequencies analytically \cite{P.Deglmann:CPL.2002}, and a normal
mode analysis, using internal coordinates, shows that the lowest
frequency vibration corresponds almost purely to the torsion of the
phenyl rings. We compute vibrational Raman cross-sections from
derivatives of the polarizability tensor with respect to the normal
modes of vibration. Table~\ref{table:raman} shows the frequencies
and Raman cross sections of the lowest vibrations of DFDBrBPh,
computed at B3LYP/TZVPP level. The results provide the justification
for assuming that the kick pulse primarily interacts with the lowest
normal mode, since its Raman scattering cross section is over four
times higher than the second lowest normal mode and 20 times higher
than any other normal mode. We note that the character of the second
lowest (and second strongest Raman active) mode contains no
torsional motion.
\begin{table}
\centering \caption{\label{table:raman}Vibrational frequencies and
relative Raman cross sections of the eight lowest lying normal modes
of DFDBrBPh, computed at B3LYP/TZVPP level.}
\begin{ruledtabular}
\begin{tabular}{l|rr}
mode & frequency (cm$^{-1}$) &  cross section ($\%$) \\
\hline
$\nu_1$ &  23.9      &  100.0 \\
$\nu_2$ &  55.7      &   21.4 \\
$\nu_3$ &  60.7      &    2.8 \\
$\nu_4$ & 111.9      &    5.1 \\
$\nu_5$ & 145.2      &    1.3 \\
$\nu_6$ & 194.7      &    3.6 \\
$\nu_7$ & 195.3      &    1.3 \\
$\nu_8$ & 213.4      &    0.8 \\
\end{tabular}
\end{ruledtabular}
\end{table}

\section{Experimental methods}
\subsection{Synthesis of 3,5-dibromo-3',5'-difluorobiphenyl}
\begin{figure}
\includegraphics[width=\columnwidth]{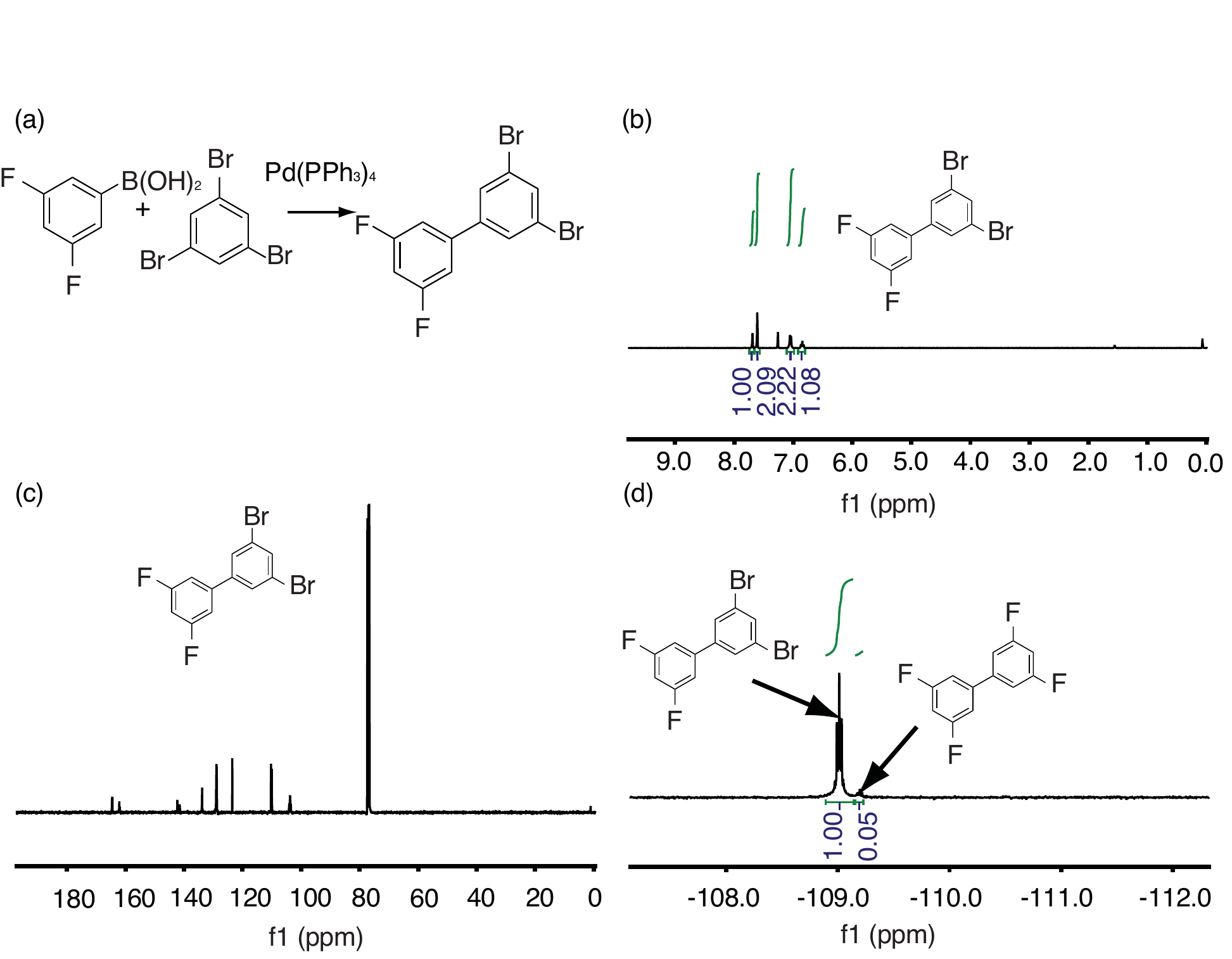}
\caption{\label{fig4} (Color online) (a), Synthesis of DFDBrBPh by
Pd-catalyzed cross-coupling. (b)-(d), NMR-spectra of synthetic
DFDBrBPh. Integrated signals are indicated by green (grey) curves, and the
corresponding numbers are situated directly below the respective
curves.}
\end{figure}

To a round-bottom flask equipped with a magnetic stirring bar under
argon we add 3,5-difluorophenyl-boronic acid (474 mg, 3.0 mmol),
1,3,5-tribromobenzene (1134 mg, 3.6 mmol), and Pd(PPh$_3$)$_4$ (69.6
mg, 2 mol$\%$). The flask is then evacuated and backfilled with
argon twice. To this mixture of solids we add toluene (18 mL,
degassed), EtOH (6 mL, degassed), and 2M aq. Na$_2$CO$_3$ (3 mL,
degassed). We next heat the mixture to 90$^\circ$C under argon for
5.5 h and cool it to room temperature. The mixture is diluted with
water and extracted three times with Et$_2$O. We dry the combined
organic fractions with MgSO$_4$, filter and concentrate to give
approximately 1.4 g of a crude mixture. Based on NMR we find that
this mixture contains both the desired mono-Suzuki coupling product
\cite{Miyaura:CR.1999}, double-coupled products, and a homo-coupling
product from the boronic acid. Careful column chromatography on
SiO$_2$ eluting with n-hexane and c-hexane finally allows us to
isolate 182 mg 3,5-dibromo-3',5'-difluorobiphenyl (DFDBrBPh) as a
white amorphous solid (contains $<$3$\%$ of the homocoupling product
DFDBrBPh). NMR spectroscopic data (see also Fig.~\ref{fig4}): $^1$H
NMR $\delta$ ($400$ MHz, CDCl$_3$) 7.69 (t, J 1.7 Hz, 1H), 7.61 (d,
J 1.7 Hz, 2H), 7.05 (m, 2H), 6.85 (m, 1H). $^{13}$C NMR $\delta$
(100 MHz, CDCl$_3$) 163.5 (2C), 142.3, 141.5, 133.8, 128.9 (2C),
123.5 (2C), 110.1 (2C), 103.8. Elemental analysis, calculated: C
41.42$\%$, H 1.74$\%$, Br 45.92$\%$; found: C 41.20$\%$, 1.66$\%$,
Br 45.90$\%$.

\subsection{Setup of the laser experiment}
\begin{figure}
\includegraphics[width=\columnwidth]{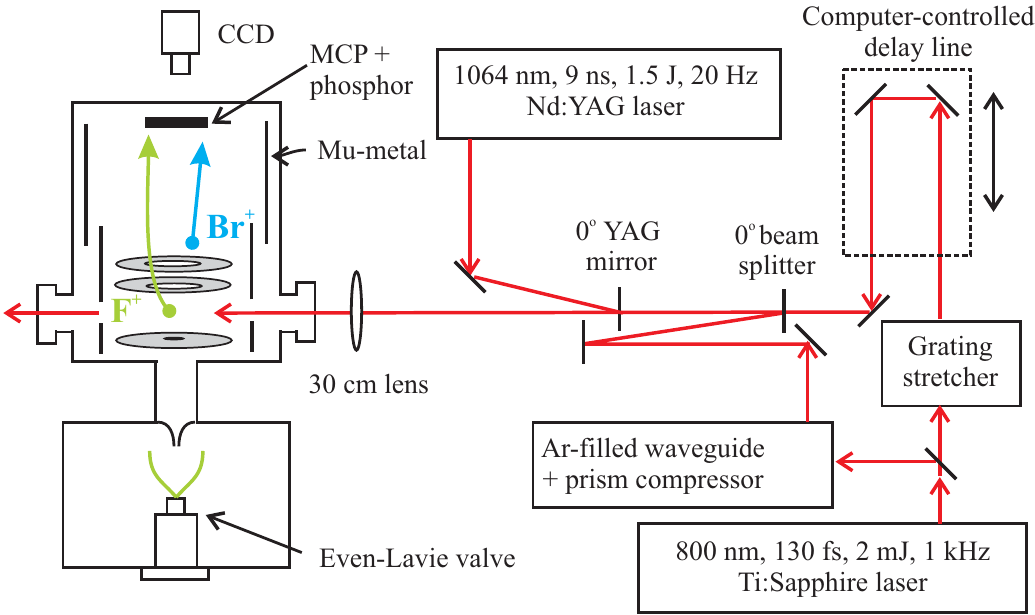}
\caption{\label{fig5}(Color online) Schematic representation of the
experimental setup, showing the vacuum system and the three laser
beam paths. The Nd:YAG (alignment) pulses and the probe pulses are
polarized vertically, i.e., perpendicular to the detector plane,
whereas the fs kick pulse is polarized in the plane of the detector.
}
\end{figure}

A schematic of the experimental setup is shown in Fig.~\ref{fig5}.
Less than hundred milligram of solid DFDBrBPh is heated to
130$^\circ$C and expanded in 90 bar of He into vacuum using a pulsed
Even-Lavie valve \cite{even:2000:jcp,kumarappan:2006:jcp}. The
resulting supersonic molecular beam is skimmed and subsequently
crossed by three, focused, pulsed laser beams at 90$^\circ$. The
first laser beam originates from a Q-switched Nd:YAG laser and
consists of 9-ns-long pulses at 1064 nm. It is focused to a spotsize
$\omega_0$ $\sim$ 35 $\mu$m yielding a peak intensity of $\sim
7\times10^{11}$ W/cm$^2$. The purpose of these pulses is to align
the stereogenic axis of the DFDBrBPh molecules. The second laser
beam consists of 0.7-ps-long kick pulses ($\lambda$ = 800 nm),
obtained by passing part of the output from an amplified Ti:Sapphire
fs laser system through a grating stretcher. The spotsize of
$\omega_0$ = 43 $\mu$m yields a peak intensity of $\sim
5\times10^{12}$ W/cm$^2$. These kick pulses are used to induce
torsion and rotation of the molecules. The third laser beam consists
of 25-fs-long pulses, which we achieve by passing another part of
the output from the Ti:Sapphire system through a hollow wave guide
compressor setup. The pulses have a spotsize $\omega_0$ = 25 $\mu$m
and a peak intensity of $\sim 2\times10^{14}$ W/cm$^2$. These probe
pulses Coulomb explode the molecules and the produced ions are
extracted by a weak static electric field, in a velocity imaging
geometry, and projected onto a two-dimensional detector, consisting
of a micro channel plate (MCP) detector backed by a phosphor screen.
We detect the F$^+$ and Br$^+$ ions separately by time gating of the
MCP detector around the respective arrival times of the ions. We
record the ion images on the phosphor screen by a CCD camera and
determine the coordinates of each individual ion hit.
Figure~\ref{fig6}(a) displays the ion images of F$^+$ and Br$^+$ at
selected probe times.

\section{Results and Discussion}\label{sec:results}
\begin{figure}
    \centering
    \includegraphics[width = \columnwidth]{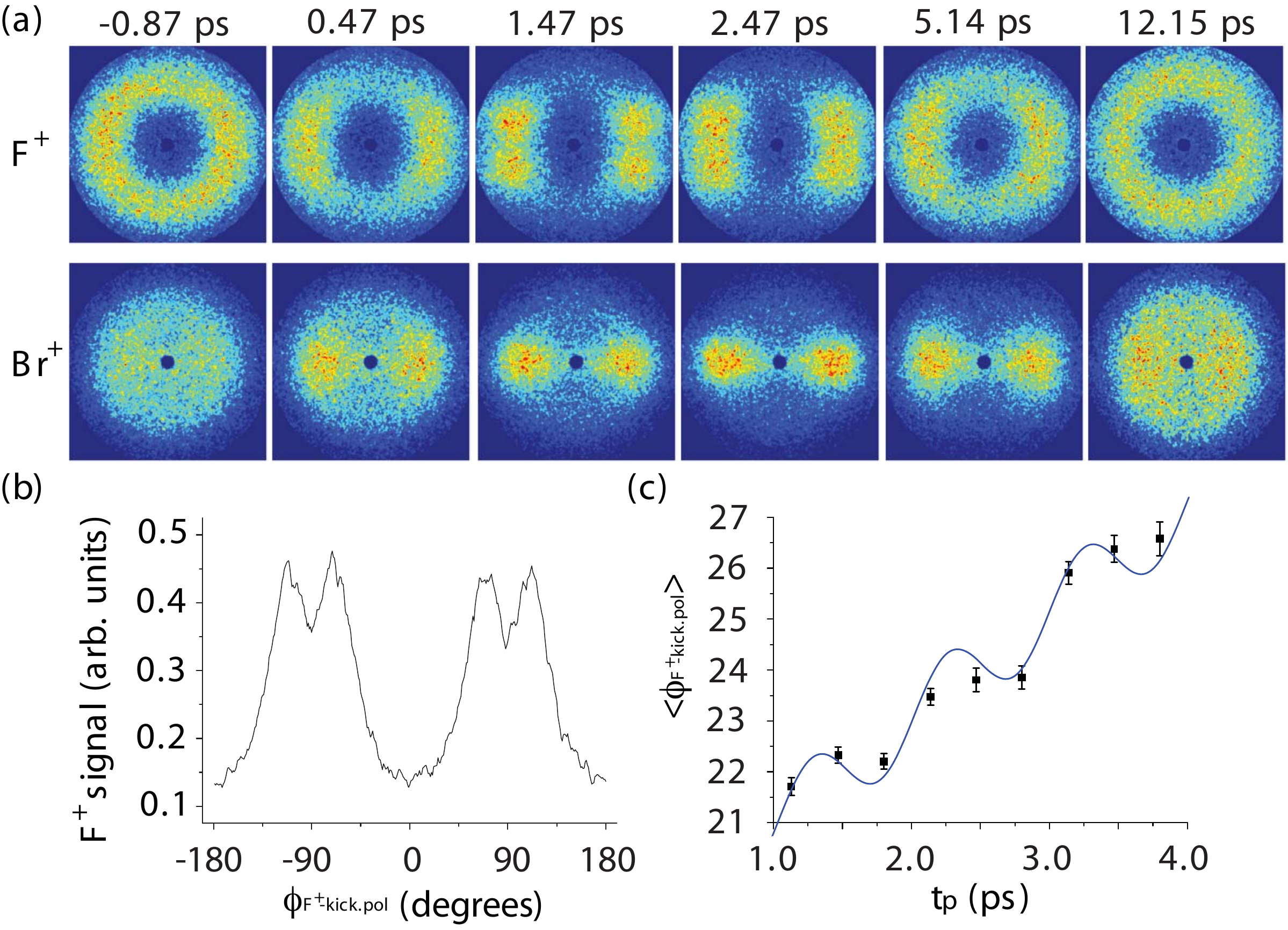}
 \caption{
 (Color online)
 (a)  Ion images of F$^+$ and Br$^+$ fragments at probe times $t_p$.
 The ns pulse is polarized perpendicularly to the image (detector)
plane and the  $5\times10^{12}$ W$/$cm$^2$,
 $0.7$ ps (FWHM) kick pulse is polarized horizontally. (b) Angular
distribution of the F$^+$ ions, at $t_p$~=1.47 ps, obtained by
radially integrating the corresponding F$^+$ ion image. The
splitting of the pairwise peaks is twice the average angle,
$\langle\phi_{\text F^\text+ \text{- kick. pol}}\rangle$, between
the F$^+$ ion recoil (and thus the F-phenyl plane) and the kick
pulse polarization. (c) $\langle\phi_{\text F^\text+ \text{- kick.
pol}}\rangle$ as a function of $t_p$, for times where a clear
four-peak structure is visible in the angular distributions. The
curve is a fit of the sum of a linear and a harmonic function to the
experimental points (squares). }
 \label{fig6}
\end{figure}

First, we establish that the ns pulse is able to hold the C-C bond
axis of DFDBrBPh along its polarization direction. At $t_p = -0.87$
ps the F$^+$ image, displayed in Fig.~\ref{fig6}(a), is almost
circularly symmetric and the small deviation from circular symmetry
is explained by noting that the kick pulse has a finite value at
$-0.87$ ps. The absence of ions in the innermost region is only
compatible with the C-C bond axis being aligned perpendicular to the
detector plane and the F-phenyl planes uniformly
distributed around the C-C bond axis. The corresponding Br$^+$ image
is also essentially circularly symmetric although the hollow-out of
the center is not as pronounced as for F$^+$.

Next, we investigate the effect of the kick pulse. Already at $t_p =
0.47$ ps the circular symmetry is broken and both F$^+$ and Br$^+$
ions start to localize around the polarization direction of the kick
pulse. The F$^+$ ions remain within the radial range of the ring
structure, which shows that the kick pulse does not cause any
significant distortion of the C-C bond axis alignment but rather
initiates an overall rotation of the molecule around this axis. The
latter is expected since the torque imparted by the kick pulse will
force the SMPA (Fig.~2) to align along the kick pulse polarization
after a delay determined by the kick strength. At $t_p = 1.47$ ps
the localization sharpens for both ions species, but whereas the
Br$^+$ ions are simply more localized around the polarization axis
the F$^+$ ion distribution exhibits a four dot structure. This
behaviour is compatible with alignment of the SMPA along the kick
pulse polarization. In the limit of perfect SMPA alignment, the
Br$^+$ ions would appear as two pairs of strongly localized regions
on the detector corresponding to molecules with their Br-phenyl
plane located either 11$^\circ$ clockwise or counterclockwise to the
SMPA. Each pair of Br$^+$ ions would then be angularly separated by
22$^\circ$. In practise and consistent with theory (see
Fig.~\ref{fig8}) the SMPA alignment is not strong enough to resolve
the two Br$^+$ regions. It is, however, sharp enough to resolve the
two pairs of F$^+$ ion hits due to the much larger offset
(28$^\circ$) of the F-phenyl plane from the SMPA. At $t_p = 2.47$ ps
the Br$^+$ distribution has localized further showing that the
Br-phenyl plane has rotated into stronger alignment with the kick
pulse polarization. If the torsion was unaffected, i.e., $\phi_d$
remained unchanged, the F$^+$ ion image at $t_p = 2.47$ ps should
exhibit a distinct four-dot structure similar to the image at $t_p =
1.47$ ps but with a larger angular splitting between the ion regions
in each of the two pairs. Clearly, the four-dot structure at 2.47 ps
is significantly blurred compared to the case at 1.47 ps. Thus, we
conclude that the kick pulse not only sets the molecule into
controlled rotation around the C-C axis, it also initiates torsional
motion.

Beyond 3.8 ps the angular confinement of both the F$^+$ and the
Br$^+$ ions is gradually lost due to continued rotation around the
C-C axis with dihedral dynamics imposed, and the images for both ion
species eventually regain their circularly symmetric form (see
images at 12.15 ps in Fig.~\ref{fig6}(a)) identical to the pre-kick
pulse images at -0.87 ps.

Further insight into the effect of the kick pulse is obtained by
analyzing the angular distribution of the F$^+$ ions as a function
of $t_p$. An example of the F$^+$ angular distribution at $t_p$ =
1.47 ps is shown in Fig.~\ref{fig6}(b). Figure~\ref{fig6}(c)
displays the average angle between the F-phenyl rings and the kick
pulse polarization as a function of $t_p$ in the time interval where
a clear four-peak structure is visible in the angular distributions.
The increase from 22.5$^\circ$ at 1.47 ps to 26.5$^\circ$ at 3.8 ps
shows that the F-phenyl plane gradually moves away from the kick
pulse polarization due to the overall rotation of the molecule
around the C-C bond axis, and we ascribe the recurrent dips to a
periodical variation in $\phi_d$. We estimate the period to be $\sim
1$ ps and the amplitude to $\sim 0.6^\circ$ for this oscillation.

\begin{figure}
    \centering
    \includegraphics[width=\columnwidth]{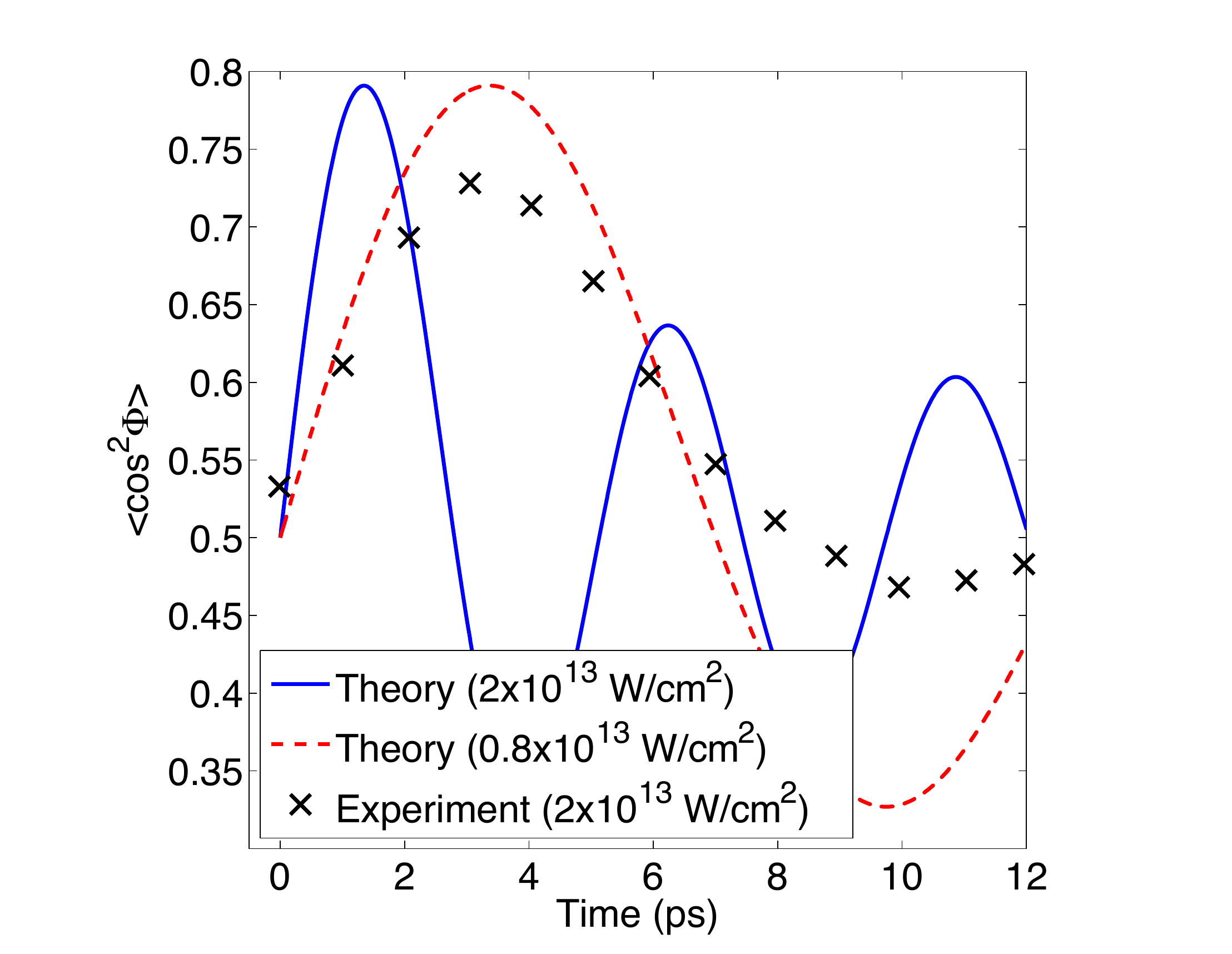}
 \caption{(Color online) Comparison of the classical model for rotation at various intensities with the experiment of Ref.~\cite{Viftrup:prl:2007}. The theory captures the laser induced rotation within the first 12 ps, but generally overestimates the degree of angular confinement.}
  \label{fig7}
\end{figure}

\begin{figure}
    \centering
    \includegraphics[width=\columnwidth]{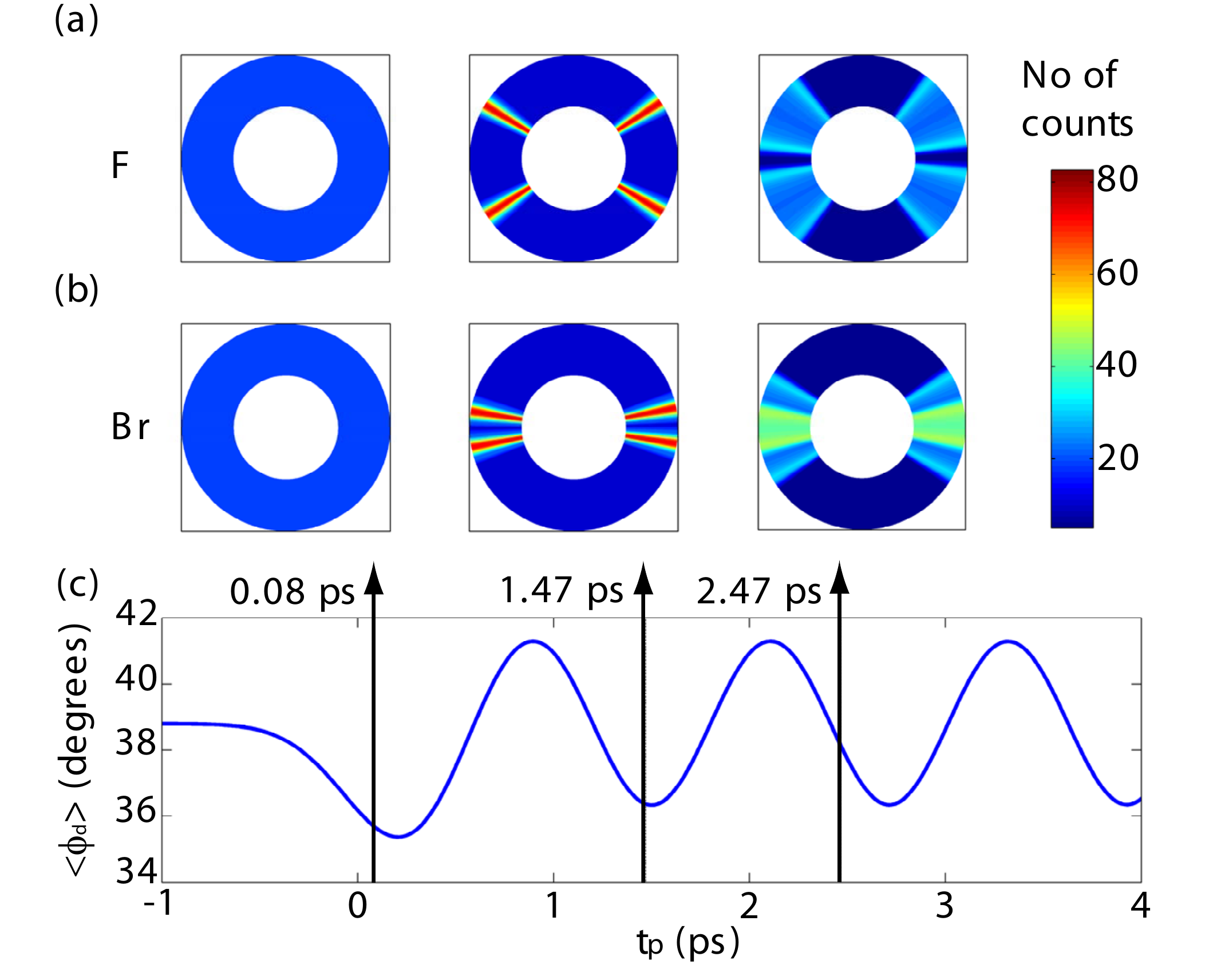}
 \caption{(Color online) Angular distributions of (a) F-phenyl and (b) Br-phenyl rings at $t_p=0.08$, 1.47 and 2.47 ps. (c) Expectation value of the dihedral angle for a
molecule starting out with the SMPA aligned along the kick pulse
polarization. The kick pulse is as in Fig.~\ref{fig6}.
 }
  \label{fig8}
\end{figure}

Before presenting the theoretical results on DFDBrBPh, we want to
confirm the validity of the classical treatment of rotation. To this
end we apply the model to a test case, where the simpler
3,5-diflouroiodobenzene (DFIB) has its most polarizable axis fixed
by a ns laser pulse and is then set into rotation about this axis by
an orthogonally polarized femtosecond pulse. Here, there is no
torsion and the rotational dynamics have been observed directly
recently~\cite{Viftrup:prl:2007}. In Fig.~\ref{fig7}, we compare the
theoretical and experimental results and find good agreement.

Figure~\ref{fig8} shows the results of a calculation with laser
parameters identical to the experimental values and an initial
rotational temperature of 0 K. Prior to the kick pulse the angular
distributions of the Br- and F-phenyl rings are isotropic as in the
experiment. Maximum alignment of the SMPA occurs at 1.3 ps and the
confinement of the F-phenyl rings at a large angle with respect to
the kick pulse polarization (cf.~middle panel, Fig.~\ref{fig8}(a))
explains the distinct four-dot structure observed at $t_p = 1.47$ ps
in the experimental F$^+$ ion image (Fig.~\ref{fig6}(a)). Also, at
$t_p = 1.47$ ps the confinement of the Br-phenyl rings at a small
angle with respect to the kick pulse polarization predicts a much
less distinct, if any, four-dot structure in good agreement with the
Br$^+$ ion image.  At $t_p = 2.47$ ps the angular localization of
the F-phenyl rings has broadened (right panel, Fig.~\ref{fig8}(a))
and a blurred four-dot structure is seen, consistent with the
experimental result at $t_p = 2.47$ ps. The distribution of the
Br-phenyl rings is also broadened (right panel, Fig.~\ref{fig8}(b)),
but remains localized around the kick pulse polarization fully
consistent with the Br$^+$ ion distribution, recorded at 2.47 ps.

The theoretical value $\langle \phi_d \rangle$ exhibits oscillations
(Fig.~\ref{fig8}(c)) with a period of $\sim1.2$ ps and amplitude of
$\sim 2.45^\circ$. The period agrees well with the experimental
value ($\sim 1$ ps), and smaller modulation in
$\langle\phi_d\rangle$ is expected in the experiment ($\sim
0.6^\circ$) partly since here the SMPA is not pre-aligned and partly
due to deviations from the stated laser peak intensity. The behavior is ascribed
to a wave packet of normal modes in the torsional double well
potential (Fig.~\ref{fig3}(b)) for a molecule starting out with the
SMPA aligned. The qualitative agreement of Figs.~\ref{fig6}(b) and
\ref{fig8}(c) corroborates the interpretation of the kick pulse
inducing time-dependent torsional motion.

\section{Perspectives}
We will now discuss some perspectives of the demonstrated laser
control of torsion. For one thing, we suggest a time-resolved study
of
de-racemization~\cite{Faber:2001,Fujimura:cpl:1999,Shapiro:prl:2000,kroner:cp:2007,kroner:pccp:2007},
where one enantiomer is steered into its mirror form. To do this we
break the inversion symmetry of the C-C bond axis by orienting each
molecule. This allows us to discriminate between the two
enantiomeric forms.
Theory~\cite{friedrich:1999:jpca,friedrich:1999:jcp} and
experiment~\cite{tanji:pra:2005,holmegaard:2009:prl} show that
orientation can be added to 3D alignment by combining the ns
alignment pulse with a static electric field. Next, the excitation
of torsional motion is optimized by eliminating initial rotation
around the C-C bond through alignment of the SMPA (Fig.~2) prior to
the kick pulse using an elliptically rather than a linearly
polarized ns pulse \cite{stapelfeldt:2003:rmp,Seideman:prl:2007}.
Finally, the interaction strength between the molecule and the kick
pulse needs to be increased either through higher intensity, a
longer kick pulse or by trains of synchronized kick
pulses~\cite{leibscher:2003:prl,viftrup:pra:2006,lee:jpb:2004}.
\begin{figure}
    \centering
    \includegraphics[width=\columnwidth]{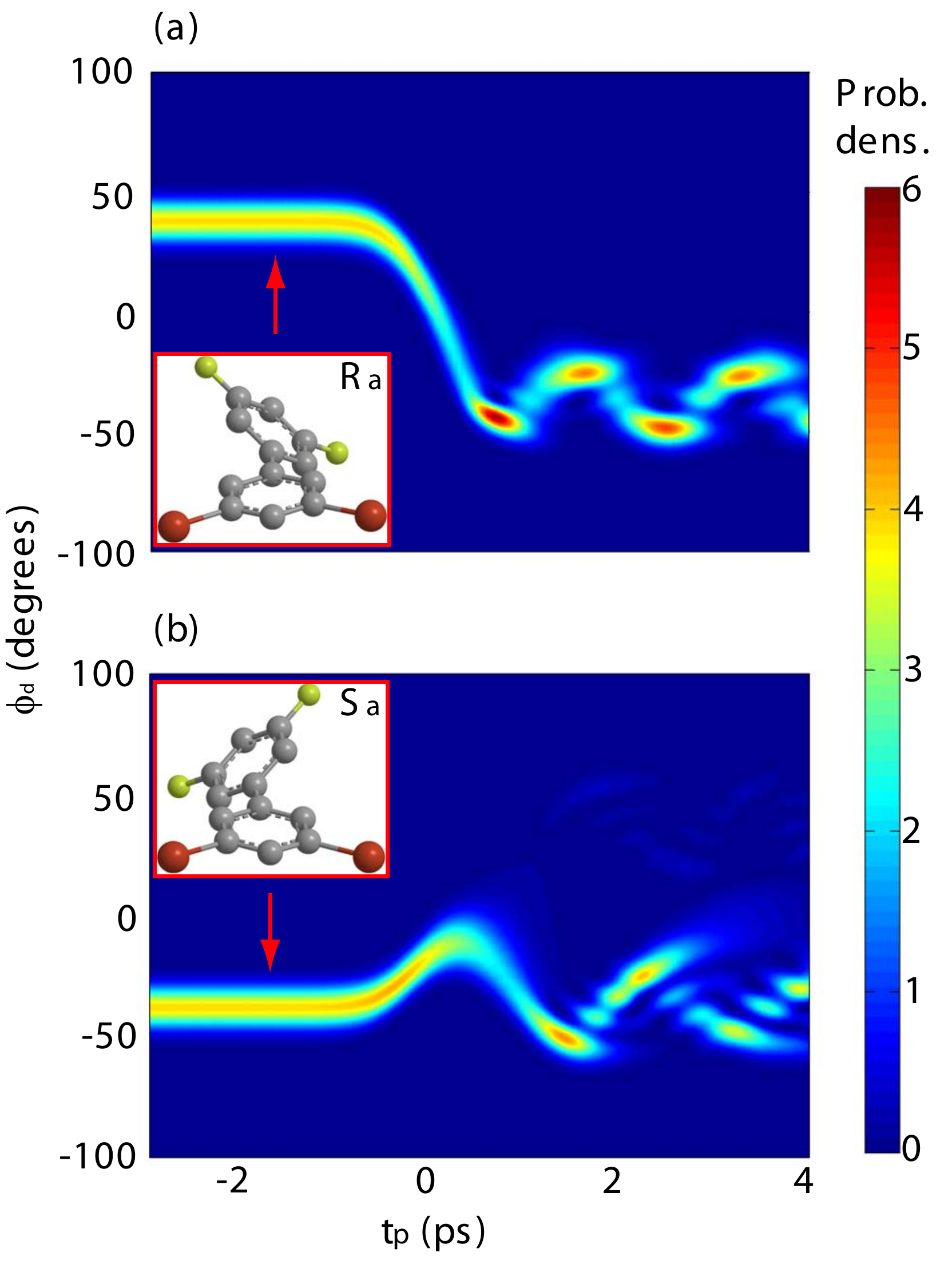}
\caption{ (Color online) Time evolution of the dihedral angle for a
molecule starting out as (a) an R$_a$ or  (b) an S$_a$ enantiomer.
Initially, the molecule is 3D oriented with the Br-phenyl end
pointing out of the paper and the SMPA aligned at an angle of
13$^\circ$ with respect to the kick pulse polarization. The kick
pulse triggering the torsional motion has a peak intensity of
$1.2\times10^{13}$ W/cm$^2$ and duration (FWHM) of $1.0$ ps. The
torsional potential is scaled down by 1/4 rather than by increasing
the kick strength. In practise reduction of the torsional barrier is
possible, for instance, by modifying the aromatic rings or by using
halogen substituted biphenylacetylene.}
  \label{fig9}
\end{figure}

Assuming initial orientation and confinement of the SMPA, we have
calculated the $\phi_d$ dynamics for both conformations the DFDBrBPh
molecule with a reduced torsional barrier (see caption to Fig.~\ref{fig9}). The results are shown in Fig.~\ref{fig9}, and clearly,
the present method would allow for a time-resolved study of the
transition from one enantiomer into the other. A quantitative
analysis of the efficiency of the process shows that after the pulse
$99\%$ of the molecules starting out as R$_a$ has changed into S$_a$
enantiomers, whereas only $13\%$ of the S$_a$ enantiomers changed
into R$_a$. The inverse process causing an excess of R$_a$
enantiomers, is achieved simply by inverting the orientation of the
molecules.

We believe that the results presented in this paper also fuel the field of molecular junctions~\cite{Ratner:science:2003,tour:science:2001,chen:cp:2002} with new possibilities. The conductivity of molecules like biphenyl will depend on the dihedral angle~\cite{seminario:JACS:1998}, and as such this type of molecules, if placed between two leads, can function as a molecular junction~\cite{cizek:PRB:2004,cizek:czechjphys:2005,benesch:cpl:2006,benesch:JPhysChem:2008} that may be used to control the charge flow and act as, e.g., a switch~\cite{Seideman:prl:2007}. Importantly, we can realize quantum control of such a system by means of kick pulse schemes and in this way tailor very specific torsional wave packets~\cite{Werschnik:JPB:2007} to dictate the detailed time-dependent dihedral motion and thereby the current flow through the molecular junction. Seen through the light of controllability, molecular junctions based on laser controlled torsion complements the available schemes, such as the mechanical break junctions~\cite{Wu} and the previously suggested resonantly light driven molecular junctions~\cite{Loudwig:JACS:2006,Zhang:PRL:2004,valle:naturen:2007}.


\section{Conclusion}
We presented a fs time-resolved study on laser controlled torsion of
axially chiral molecules. Experimentally, the symmetry axis of
DFDBrBPh molecules was held fixed in space by a long alignment laser
pulse while a fs kick pulse, polarized perpendicular to the fixed
axis, was applied. The torsion as well as the overall rotation of
the molecule was monitored with fs time resolution using a delayed
probe pulse. To substantiate the observations, we developed an
original theoretical model, which produces results in good agreement
with the experiment. The model divides the dynamics of the molecule
into a the rotation with classical behavior and quantum mechanical
torsional dynamics. As such the model provides a transparent
physical intepretation of essential features of laser controlled
torsion of DFDBrBPh. Based on the theoretical model, we further
discussed the extension of the experiment to a setup that will
facilitate a fs time-resolved study of de-racemization. Finally, we
pointed to perspectives of using molecules like DFDBrBPh in
molecular junctions in order to realize laser controlled charge flow.

\section{Acknowledgements}
This work was supported by the Carlsberg Foundation, The Lundbeck
Foundation, Danish Natural Research Foundation, the Danish Natural
Science Research Council, and the Danish Research Agency (Grant no.
217-05-0081).

\end{document}